\documentclass[aps,prb,twocolumn,groupedaddress]{revtex4}
\usepackage{amsfonts}
\usepackage{amssymb}
\usepackage{amsmath}
\usepackage{graphicx}

\usepackage{dcolumn}   
\usepackage{bm}        
\usepackage{flafter}

\begin{document}


\title{Effect of the spin-orbit interaction on
the thermodynamic properties of  crystals:\\ The specific heat of
bismuth}

\date{\today}

\author{D\rm\'{\i}az-Luis}
\affiliation{CINVESTAV-Quer\'{e}taro, Libramiento Norponiente 2000,
Fraccionamiento Real de Juriquilla 76230 Quer\'{e}taro, Qro.
Mexico\\
and Unit\'{e} de Physico-Chimie et de Physique des
Mat\'{e}riaux Universit\'{e} Catholique de Louvain\\
B-1348 Louvain-la-Neuve, Belgium}
\author{A.H. Romero}
\affiliation{CINVESTAV-Quer\'{e}taro, Libramiento Norponiente 2000,
Fraccionamiento Real de Juriquilla 76230 Quer\'{e}taro, Qro. Mexico}
\author{M. Cardona}
\affiliation{Max-Planck-Institut f\"{u}r Festk\"{o}rperforschung,
Heisenbergstra$\rm\beta$e 1, D-70569 Stuttgart, Germany}
\author{R. K. Kremer}
\affiliation{Max-Planck-Institut f\"{u}r Festk\"{o}rperforschung,
Heisenbergstra$\rm\beta$e 1, D-70569 Stuttgart, Germany}
\author{X. Gonze}
\affiliation{Unit\'{e} de Physico-Chimie et de Physique des
Mat\'{e}riaux Universit\'{e} Catholique de Louvain\\
B-1348 Louvain-la-Neuve, Belgium }

\begin{abstract}
In recent years, there has been increasing interest in the specific
heat $C$ of insulators and semiconductors  because of the
availability of samples with different isotopic masses and the
possibility of performing \textit{ab initio} calculations of its
temperature dependence $C(T)$ using as a starting point the
electronic band structure. Most of the crystals investigated are
elemental (e.g., germanium) or binary (e.g., gallium nitride)
semiconductors. The initial electronic calculations were performed
in the local density approximation and did not include spin-orbit
interaction. Agreement between experimental and calculated results
was usually found to be good, except for crystals containing heavy
atoms (e.g., PbS) for which discrepancies of the order of 20\%
existed at the low temperature maximum found for $C/T^3$. It has
been conjectured that this discrepancies result from the neglect of
spin-orbit interaction which is large for heavy atoms
($\Delta_0\sim$1.3eV for the $p$ valence electrons of atomic lead).
Here we discuss measurements and \textit{ab initio} calculations of
$C(T)$ for crystalline bismuth ($\Delta_0\sim$1.7 eV), strictly
speaking a semimetal but in the temperature region accessible to us
($T >$ 2K) acting as a semiconductor. We extend experimental data
available in the literature and notice that the \textit{ab initio}
calculations without spin-orbit interaction exhibit a maximum at
$\sim$8K, about 20\% lower than the measured one. Inclusion of
spin-orbit interaction decreases the discrepancy markedly: The
maximum of $C(T)$ is now only 7\% larger than the measured one.
Exact agreement is obtained if the spin-orbit hamiltonian is reduced
by a factor of $\sim$0.8.

\end{abstract}
\smallskip

\pacs{N.N.}

\maketitle
\section{Introduction}
In the past few years a number of investigations on the dependence
of the specific heat of semiconductors and insulators on temperature
and isotopic masses have been carried out. These works involved
careful low temperature experiments (for elemental crystals see
Refs. \onlinecite{Cardona0} and \onlinecite{Gibin}, for binaries see
Refs. \onlinecite{Serrano} and references therein) and elaborate
\textit{ab initio} calculations based on electronic band structures
computed in the framework of the local density approximation (LDA).
Recent work involving binary lead chalcogenides with different
isotopic compositions (PbS, (Ref. \onlinecite{Cardona1}), PbSe (Ref.
\onlinecite{Cardona2})) yields a low temperature maximum in the
quantity $C(T)/T^3$ (at $\sim$12K)\cite{Cardona1,Cardona2} about
25\% lower than the measured one. Correspondingly, the calculated
phonon dispersion relations are on the average 10\% higher than
those determined with inelastic neutron scattering
(INS)\cite{Elcombe}.The higher phonon frequencies qualitatively
explain the lower calculated specific heats.

When trying to ascertain whether the \textit{harder} phonon
frequencies (and lower specific heats) calculated for PbS and PbSe
were due to the lack of spin-orbit (\textit{s-o}) coupling in the
\textit{ab initio} electronic structure some difficulties arose.
These involved not only the extended computational time but also
divergences in the dispersion relations of the optical phonons for
\textbf{\textit{q}}$\rightarrow$0. The latter may be related to the
strongly ionic, near ferroelectric character of these compounds.

To our knowledge the computation of the
\textbf{\textit{q}}$\rightarrow$0 divergences when the spin-orbit
coupling is present has not yet been implemented in existing
first-principle software. By contrast, for metals, there is no such
divergence, which makes possible the direct computation of phonon
band structure and thermodynamical properties based on
Density-Functional Perturbation Theory (Ref.
\onlinecite{Baroni,Lee,ABINIT1}), including spin-orbit, as
implemented in the ABINIT software\cite{ABINIT}. Thus, we performed
measurements and calculations for crystalline bismuth, which is free
of the ionic divergences present in the lead chalchogenides.

In this Letter we demonstrate that inclusion of \textit{s-o}
coupling, in fact, considerably reduces the discrepancies between
experimental heat capacity data and \textit{ab initio} results. In
addition, we also discuss the dependence of the rhombohedral lattice
parameter $a_0$ on the magnitude of \textit{s-o} coupling.

Bismuth is a semimetal closely related to the lead chalcogenides: It
has 10 valence electrons per primitive cell and a rhombohedral
structure which can be derived from that of PbS by making both atoms
equal and applying a Peierls-like distortion  to the PbS cube,
involving an elongation of one of its [111] axes \cite{Bellisent}.
The distortion converts the simple cubic structure of bismuth, with
one atom per primitive cell, into a rhombohedral one with two atoms
per primitive cell (two sublattices). This structure is
characterized by three parameters: the bond length $a_0$, the
rhombohedral angle and a shift between the two sublattices
\cite{Murray,Diaz}. These parameters were determined by energy
minimization using the ABINIT code. The results obtained for these
parameters in Refs. \onlinecite{Murray} and \onlinecite{Diaz} with
and without \textit{s-o} coupling differ by less than 1\%.

The phonon dispersion relations were calculated with ABINIT, in Ref.
\onlinecite{Murray} with \textit{s-o} coupling whereas in Ref.
\onlinecite{Diaz} calculations with and without \textit{s-o}
coupling were performed. With \textit{s-o} coupling, excellent
agreement with experimental results was obtained whereas without it,
discrepancies of the order of 10\% were found, the calculated bands
lying higher in frequency than the measured ones. Similar results
were found for PbS (Ref. \onlinecite{Cardona1})  and PbSe (Ref.
\onlinecite{Cardona2}) without \textit{s-o} coupling. In view of
these results we proceeded to calculate the specific heat $C(T)$ of
bismuth. Measurements which  were performed on high purity
(99.9999\%, Preussag Pure Metals) single crystals in order to
complete the available experimental data (performed for
polycrystalline samples)\cite{Touloukian,Keesom,Franzosini,
Cetas,Archer}. Recent data collected by  various earlier authors and
our own data (the experimental technique is described in
Refs.\onlinecite{Cardona0,Gibin,Serrano,Cardona1}) are compiled in
Fig. \ref{fig1}, together with the results of our \textit{ab initio}
calculations performed with and without spin-orbit coupling.

\section{Results and Discussion}
\begin{figure}[h]
\includegraphics[width=8.5cm]{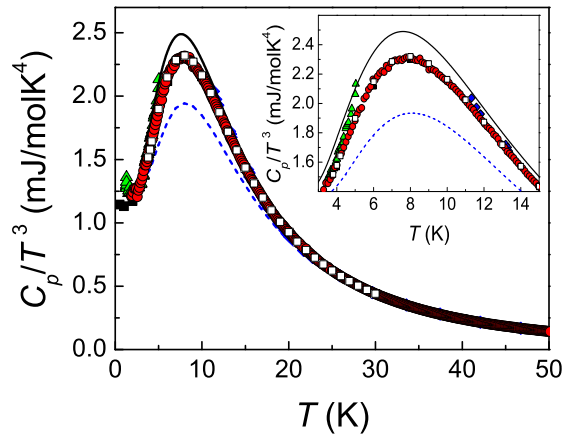}
\vspace{-3mm} \caption{\label{fig1}(Color online) Heat capacity of a
Bi single crystal, purity 99.9999\%. (red filled) circles as
measured in this work compared with literature data obtained on
polycrystalline samples. (green) $\blacktriangle$ (Ref.
\onlinecite{Keesom}); (blue) {\footnotesize{$\blacklozenge$}} (Ref.
\onlinecite{Franzosini}); $\Box$ (Ref. \onlinecite{Cetas}); (black)
{\tiny{$\blacksquare$}} (Ref.  \onlinecite{Archer}). (black) solid
line: ABINIT results with spin-orbit coupling included; (blue)
dashed line: ABINIT calculation without spin-orbit coupling.}
\end{figure}

The hitherto available experimental points were measured on
polycrystalline samples. They are rather widely spaced in
temperature, with the exception of Keesom's below 4 K
(Ref.\onlinecite{Keesom,Phillips}). We have therefore performed
measurements of $C_p(T)$ on single crystals for $T$ between 1.8 and
100 K with the measuring temperatures spaced by $\sim$0.1K up to 50K
and 0.5K steps above 50K. The maximum of $C(T)/T^3$ takes place at
7.5 K. According to Ref. \onlinecite{Cardona2} it should be found at
$\sim T_{TA}$/6, where $T_{TA} \sim$42 K is an Einstein oscillator
frequency which can be read off the phonon density of states
\cite{Murray}. The temperature of the maximum in Fig. \ref{fig1} is
found to be 7.5 K, fairly close to $T_{TA}$/6 = 7 K.

Our measurements place the maximum value of $C/T^3$ at 2.320 $\pm$
0.03 mJ/mol K$^4$. The value calculated without \textit{s-o}
splitting lies at 1.940 mJ/mol K$^4$ whereas with \textit{s-o}
coupling one finds 2.500 mJ/mol K$^4$. Hence, the calculation
without \textit{s-o} coupling lies 27\% below the experimental data,
that with \textit{s-o} coupling only 7\% above. Not only is the
difference between measured and calculated values of $C/T^3$ three
times smaller when \textit{s-o} interaction is taken into account,
but the agreement with \textit{s-o} coupling above 12 K lies within
the experimental error, whereas without \textit{s-o} coupling a
considerable difference is found.

\begin{figure}[h]
\includegraphics[width=8.5cm]{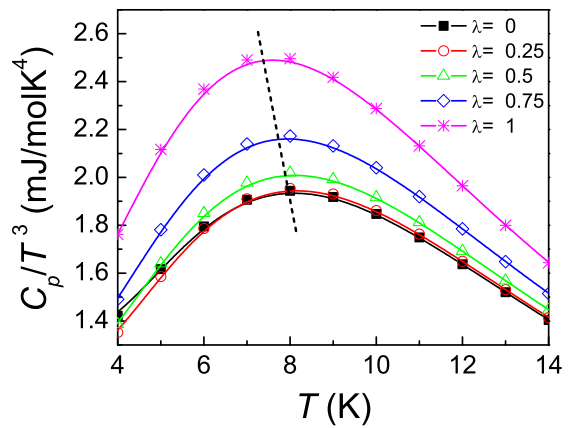}
\vspace{-3mm} \caption{\label{fig2}(Color online)Calculated heat
capacities of Bi with varying magnitude of the \textit{s-o} coupling
as indicated in the inset. Note that the maximum shifts to lower
temperatures (inclined dashed line) with increasing \textit{s-o}
coupling, as corresponds to decreasing phonon frequencies (see
text).}
\end{figure}

This improvement illustrates the importance of \textit{s-o}
interaction for the calculations of thermodynamic properties
starting from \textit{ab initio} electronic band structures for
systems containing heavy atoms and confirms our conjecture that
similar discrepancies between calculated and measured values of
$C/T^3$ found for PbS and PbSe\cite{Cardona1,Cardona2} are also due
to the lack of \textit{s-o} coupling in the electronic structure
calculations. These results suggest performing calculations and
measurements for antimony (also a semimetal with the same crystal
structure as Bi) which has a considerably smaller atomic
\textit{s-o} coupling than bismuth (0.84\,\,vs.\,\,1.7eV).

While the Sb work is in progress we have pursued yet another avenue:
we have multiplied the spin-orbit coupling Hamiltonian by a factor
$0 < \lambda <$ 1 and repeated the full \textit{ab initio}
calculations of $C(T)$ for several values  of $\lambda$. The results
obtained for bismuth are shown in Fig. \ref{fig2}.

\begin{figure}[h]
\includegraphics[width=8.5cm]{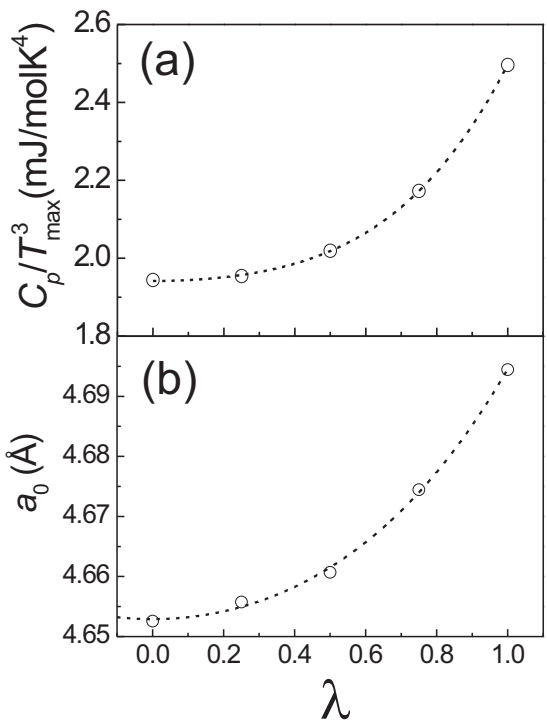}
\vspace{-3mm} \caption{\label{fig3}(a) Maxima of the quantity
$C_p(T_{max})/T_{max}$ vs. spin-orbit coupling parameter $\lambda$.
$\circ$ our data, dashed line: fit with eq. \ref{eq1} with
parameters given in the text. (b) Energy minimized lattice parameter
$a_0$ vs. spin-orbit coupling parameter $\lambda$. $\circ$ our data,
dashed line: fit with eq. \ref{eq1} with parameters given in the
text.}
\end{figure}

This figure clearly reveals the strongly supralinear dependence of
the \textit{s-o} effect on $C/T^3$ which can bet fitted with the
expression

\begin{equation}
C/T^3 = c_0 [1 + c_2\,\lambda^2\,(1 + c_4\,\lambda^2)] \label{eq1}
\end{equation}

with $c_0$=1.942(2)\,mJ/molK$^4$, $c_2$=0.116(7) and $c_4$=1.46(13).

The fit is displayed in  Fig.\ref{fig3}(a). For $\lambda$=1 the
fourth order term surpasses the quadratic term. Exact agreement
between the measured and the calculated values of $C/T^3$ is
obtained if one reduces the \textit{s-o} coupling by the factor
$\lambda$=0.8.

It is expected that other mechanical and thermodynamic properties of
Bi should depend on the \textit{s-o} coupling as well, i. e., on the
value of $\lambda$. The simplest of them is probably the
rhombohedral lattice parameter $a_0$. It is shown in
Fig.\ref{fig3}(b) calculated for the same values of $\lambda$ as
$C(T)$ and fitted with an expression similar to eq. \ref{eq1} with
$a_0(\lambda=0)\equiv c_0$=4.6529(7)\AA, $a_2$=0.0069(9) and
$a_4$=0.29(16). The values of the fit parameters $c_i$ indicate that
there is no simple relationship between the effect of \textit{s-o}
splitting on $C(T)$ and the lattice parameter $a_0(\lambda)$. The
effect of spin-orbit interaction on $C(T)$, as described by $c_2$
and $c_4$, is much larger than that on the lattice parameter
$a_0(\lambda)$. For $\lambda$=1, however, the calculated value for
$a_0$ (4.694 \AA) is also much closer to the the experimental one
(4.720 \AA) than that obtained for $\lambda$=0 (4.652 \AA).

In conclusion, we have investigated the effect of \textit{s-o}
interaction on two thermodynamic properties, $C(T)$ and $a_0$ of a
solid consisting of the heavy element bismuth. Such effects,
apparently rather substantial for a first-principles calculation of
the physical properties, have not received much consideration in the
literature, so far. For a discussion of other similar phenomena see
also Refs. \onlinecite{Norrby} and \onlinecite{Pyykkoe}.

\acknowledgments The authors acknowledge useful discussion with J.
Serrano and L. Wirtz. We thank C. Busch for providing the single
crystal of Bi and G. Siegle for experimental assistance. One of the
authors (X.G.) would like to acknowledge support from the F.R.F.C.,
project N° 2.4502.05.  A.H.R. acknowledges support by CONACYT Mexico
under Grant No. J-42647-F and D.-L. under Grant 167176. We would
like also to thank the Centro Nacional de Supercomputo at IPICyT,
Mexico for allocation of computer time.

\end{document}